BB Working Paper Series: WP No. 2021-02

# Does higher capital maintenance drive up banks' cost of equity?—Evidence from Bangladesh

Md Shah Naoaj
Mir Md Moyazzem Hosen

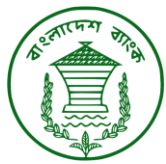

Bangladesh Bank
August 2021

# Does higher capital maintenance drive up banks' cost of equity?
## —Evidence from Bangladesh

Md Shah Naoaj
Mir Md Moyazzem Hosen[1]

## ABSTRACT

This paper assesses whether the higher capital maintenance drives up banks' cost of equity. We investigate the hypothesis using fixed-effect panel estimation with the data from a sample of 28 publicly listed commercial banks over the 2013-2019 periods. We find a significant negative relationship between banks' capital and cost of equity. Empirically our baseline estimates entail that a 10 percent increase in capital would reduce the cost of equity by 4.39 percent. We also find that the investors apparently don't price the TIER2 capital, and the perceived financial risk is heavily priced by the core capital (TIER 1 capital). Together, the results suggest that some policy adjustments—such as raising the minimum requirement of Common Equity TIER 1 capital and Additional TIER 1 capital coupled with reducing the space for Tier 2 capital—would help further strengthen banks' sustainability and financial sector stability.

---

[1] The authors work at the Banking Regulation and Policy Department of Bangladesh Bank as Deputy Director. The views, thoughts and opinions presented here are based on the authors' own analysis and do not necessarily reflect those of the authors' employer (Bangladesh Bank), other organizations or any individual or group. The authors are grateful to Dr. Md. Habibur Rahman, Executive Director (Research), and the officials of Chief Economist's Unit of Bangladesh Bank for their review and valuable comments. The authors would also like to thank Dr. Md. Deluair Hossen, Deputy Director, Bangladesh Bank for his helpful comments.

**Index**

## 1. Introduction

Ushering in the sustained macroeconomic reforms and prudent policies, Bangladesh continues to achieve robust GDP growth (8.15% growth rate in 2019), led by a fast-growing manufacturing sector and solid remittance inflows (Naoaj et al. 2021). However, like the global financial sectors, Bangladesh's financial sector is also going through internal and external shocks caused by the ongoing COVID19 pandemic. The pandemic shows that the better-capitalized banks can effectively absorb the unprecedented shocks emanating from the financial and economic sectors. Basel I, II and III standards increasingly emphasize the need for higher capital for the banks and these standards have gradually stepped up the capital requirements of banks. After the Global Financial Crisis (GFC) of 2008 in particular, Basel III standards start insisting on increasing the minimum capital requirements (MCRs) of banks to remain strong and resilient to economic and financial shocks.

However, the bank owners and the banking professionals claim that higher capital requirement increases the cost of equity, which is subsequently passed on to borrowers and, in turn, affect credit growth and dampen the real economic activity (Kashyap et al., 2010; Hanson et al. 2011).Therefore, it is an issue of debate whether the higher capital requirement results in higher cost of equity.

This research fills up the gap of empirical research on this issue in Bangladesh in line with the theory of Modigliani and Miller Hypothesis (MM Hypothesis) of 1958. The hypothesis assumes that the more equity capital a bank maintains, the less cost of equity it experiences and the more resilient it is. Therefore, this empirical research in the context of the banking sector of Bangladesh is intended to test whether the banks maintaining higher capital incur lower cost of equity.

As per MM Hypothesis and overall funding cost perspective, if the amount of equity in the capital structure increases, cost of equity declines. This is also evident from the research of Belkhir et al 2019 whereby they find that a 1-point increase in equity to asset ratio decreases the cost of equity by 18 basis points. On the other hand, for banks with low capital base, a 1-point increase in equity to asset ratio decreases the cost of equity by 79 basis points.

Though the bankers assert that the cost of equity rises with the increase in capital requirement and equity capital is indeed expensive, the idea that equity is costly is flawed from the capital regulation perspective (Admati et al. 2013). Their findings suggest that a higher amount of equity decreases the equity risk and lowers the required rate of return demanded by the equity holders. This research also suggests similar results.

In this regard, Bangladesh Bank, the central bank of Bangladesh, has given necessary directives to banks to increase the capital requirements to cope with the international best practices and to make the bank's capital more risk sensitive as well as more shock resilient. BB introduced Basel II directives in 2009 parallel to existing BRPD Circular No. 10 dated November 25, 2002. In 2014 through BRPD circular 18, BB enhanced capital requirements to 12.5% phase-in manners in line with Basel III accord. Basel III is a 2009 international regulatory accord designed to mitigate the risks of banking sector. It was formulated by the Basel Committee on Banking Supervision (BCBS) after the global financial crisis, 2008.

Table 1 summarizes the capital requirement set out in the Guidelines on Risk-Based Capital Adequacy (RBCA) by Bangladesh Bank in 2014.

Table 1: Minimum Capital Requirement

|  | 2015 | 2016 | 2017 | 2018 | 2019 |
|---|---|---|---|---|---|
| Minimum Common Equity Tier-1 Capital Ratio | 4.5% | 4.50% | 4.50% | 4.50% | 4.50% |
| Capital Conservation Buffer | - | 0.625% | 1.25% | 1.875% | 2.50% |
| Minimum CET-1 plus Capital Conservation Buffer | 4.5% | 5.125% | 5.75% | 6.375% | 7.00% |
| Minimum T-1 Capital Ratio | 5.50% | 5.50% | 6.00% | 6.00% | 6.00% |
| Minimum Total Capital Ratio | 10.00% | 10.00% | 10.00% | 10.00% | 10.00% |
| Minimum Total Capital plus Capital Conservation Buffer | 10.00% | 10.625% | 11.25% | 11.875% | 12.50% |

The central contribution of our research is that the theoretical assumptions that banks' cost of equity falls with the increase in capital requirements as per Basel III standards hold empirically correct. Though the primary motivation of this paper is based on the IMF working paper on "Bank Capital and the Cost of Equity" conducted by Mohammad Belkhir, Sami Ben Naceur, Ralph Chami, and Anis Samet at December 2019, our paper is different from Belkhir et al. 2019 in two ways—i) while they have examined the impact of capital on banks' cost of equity over a large period on 62 countries, we have examined it in the context of Bangladesh ii) while they have used the average of four different models, we use Dividend Discount Model (DDM) to identify banks' cost of equity. We use DDM as this method is best for analyzing the COE of companies that provide regular dividends, and are in matured growth stage.

## 2. Literature Review

From the perspective of society, banks' equity is not expensive and banks with better capital experience fewer changes in their lending decisions and thus perform well (Admati et al 2013). According to this research, setting capital requirements significantly higher than Basel III standards and local regulatory bodies brings higher social benefits and minimum social costs. They also observe that big financial institutions' social costs arising from increasing equity capital requirements are generally small. Therefore, we can claim that banker's assertion doesn't hold empirical testing. It is also evident that the benefits of higher capital requirements show compelling evidence for setting up minimum equity to asset ratio between 10% and 20% (Admati and Hellwig, 2014). This minimum capital requirement ensures the resilience of banks when there is any adversity in the financial sector and the economy.

Moreover, the equity holders of the banks remain safe when the debt level in the capital structure declines as the higher debt is a signal of lack of strength in the financial position of the banks. This idea is further reinforced by the research of Kashyap et al, 2010 which states that if the financial leverage of banks decreases, the equity holders' required rate of return declines and this relationship is found to be empirically true.

Though MM hypothesis asserts that markets will correctly realize the relative safety of more equity, there is still a possibility that the cost of equity may not fall due to the factors that trigger risks to the banks. Admati et al, 2013 questions the validity of such assumption and states that equity holder's required return decreases with an increase in equity in the funding mix. Kashyap et al, 2010 also finds similar results. Claims that high capital requirements result in high cost for society and bring adverse impacts on credit markets are fallacious as per their findings.

It is also argued that risk premium for equity risk declines proportionately as equity increases in the capital structure. When equity rises, the relative weight of debt in the capital mix decreases and so does the banks' riskiness. If it is assumed that equity will remain static in such situation, that would be fallacious (kashyap et al, 2010).

On the other hand, it is claimed that increased capital requirements will curb the banks' capacity to give loans to the rest of the economy, which will in turn reduce economic growth and have negative impacts on all related sectors that depend on bank financing (Admati and Hellwig, 2013). If it happens, different sectors of Bangladesh may get squeezed because of the lack of financing in terms of loan from banks. However, the cost of capital does not depend on the capital structure decision as per MM hypothesis and many economists support this view. Moreover, increased capital requirement in the banks' capital structure has several benefits such as lower equity risk, higher credit worthiness, lower leverage requirement, lower funding cost, higher resilience and shock-absorbing capacity in the long run etc. An enhanced capital requirement in the funding mix renders downside protection by reducing the equity risk and banks with higher equity capital enjoy lower systematic and idiosyncratic risk (Baker and Wurgler, 2013). These studies claim that banks with higher capital enjoy relative safety in their exposure to the market adversaries. We believe that stringent policy adopted by the regulatory body in maintaining higher capital ensures better protection for the banks. Hail and Leuz, 2009 finds that countries with greater regulation and effective enforcement systems enjoy lower capital costs. They also assert that this is also true even after considering different traditional risk factors like return volatility, financial leverage etc. and country factors such as inflation and risk-free interest rate.

Moreover, higher capital creates an overall negative drag on banks' weighted average cost of capital (WACC). WACC is the weighted average of cost of both equity and debt. The WACC declines much with the increase in capital requirement. When a bank delevers itself by increasing equity, its equity holders require reduced risk premium as deleveraging increases bank's creditworthiness and decreases the cost of borrowing. As a result, equity capital holders demand lower required return and risk premium on their invested capital.

Despite the difficulty of empirically validating this statement of lower equity risk premium as a result of deleveraging, stock returns of banks having lower debt capital tend to show lower volatility and lower beta concerning overall stock market (Modigliani and Miller, 1958). It follows that bank with lower debt capital experiences lower volatility in stock prices and this ensures the relative safety of the equity investors.

On the contrary, an increase in equity cost results from raising new equity capital and asymmetric information on new equity issuance also affects equity costs. 1% increase in new capital issuance raises the weighted average cost of capital by 2.5 basis points (Kashyap et al, 2014). However, this research also counter-intuitively suggests that high capital requirements imposed by the regulatory body motivate the banks to replace both short- and long-term debt with equity capital. Moreover, increased capital requirements benefit the banks in terms of increased systematic stability (Kashyap et al., 2010).Modigliani and Miller, 1958 also suggests that a proportionally high increase in capital requirements has a chance to bring in smaller impacts on borrowing costs of bank's clients. Moreover, the lending rate of banks will rise just only by 24-25 basis points if the minimum capital ratio is increased by 10% (Elliot, 2013).

It is argued that banks with reduced leverage experience lower stock return volatility and show lower betas. But there is no explicit link between risk measure and banks' equity. In fact, banks with lower betas tend to have higher cost of equity. Elliot, 2013 shows that bank's equity capital cost is 5% higher than the tax-adjusted cost of debt capital. However standard theory suggests that increased capital lowers the beta and subsequently the cost of banks' equity. Moreover, MM hypothesis states that equity costs remain fixed despite the variation in banks' leverage. Furthermore, higher capital in place of leverage is desirable if all other private and social costs and benefits are considered (Kashyup, Stein and Hanson, 2010).

It is also claimed that equity is expensive as dividend payments are not tax-deductible, whereas interest payments are tax-deductible. A bank may be reluctant to adjust its equity capital and may reduce its asset size to meet the regulatory requirements if equity to asset ratio is enhanced suddenly (Myers, 2001). This research warns against a rapid phase-in increase of capital requirements and simultaneously does not suggest that a gradual phase-in increase of capital will retard loan growth. If banks have sufficient time to move to the higher capital requirements using retained earnings, they will do so without reducing assets like positive NPV (Net Present Value) Loans.

Despite the robust supporters of higher bank capital requirements, the assumption that required return on equity falls with the increase in equity is doubtable (Admati et al. 2013). An empirical relationship between a rise in equity and a fall in required return on equity is extremely difficult to determine. The industry also asserts that the banks' cost of equity is not affected by the degree of leverage. But MM theorem suggests that this view of the industry is logically flawed because with a decline in bank's leverage, the riskiness of equity falls and investors demand a lower required rate of return. In fact, increasing banks' equity capital and decreasing leverage entail evidential positive results on banks' resilience, profitability and a favorable effect on the banking sector and the economy as a whole.

## 3. Methodology, Data and Variable

When a bank increases its capital maintenance, the cost of equity of the bank should reflect this increase. We test this understanding against Total Capital (Equity) Tier 1 Capital (TIER1) ratio and Capital Adequacy Ratio (CAR) by running a multivariate fixed-effect panel regression with robust standard errors. We include time and bank fixed effects in order to account for the time-constant unobserved heterogeneity. The study also employs Driscoll and Kraay (Driscoll, J. C., and A. C. Kraay1998) robust standard error model as it produces heteroskedasticity and autocorrelation-consistent standard errors that are robust to general forms of spatial and temporal dependence. The strategy of Driscoll and Kraay is to take averages of the product between independent variables and the residuals, and then to use these values in a weighted heteroskedasticity- and autocorrelation-consistent (HAC) estimator to produce standard errors (Vogelsang, 2011). For a full mathematical treatment of Driscoll Kraay standard errors, see either Driscoll and Kraay (1998) or Hoechle (2007).

In the panel model, the dependent variable is Cost of Equity (COE) and the independent variables are Total Capital (Equity) Tier 1 Capital (TIER1), Capital Adequacy Ratio (CAR), and the control variables are banks' Inefficiency (INEFF), Return on Asset (ROA), Natural Log of Total Deposit (LNDEPOSIT), Natural Log of Total Assets (LNSIZE), Asset Quality (ASSTQULTY) and Natural Log of Gross Domestic Product (LNGDP). The cost of equity is determined by using DDM following the methodology of Gordon growth model (Gordon and Shapiro 1956) to calculate cost of equity (see annex B for details). We use annual panel data set comprising cross-sectional and time-series observations and studying even when cross-sectional and time-series observations are not available. This has augmented both the quality and quantity of data set (Gujarat, 2003).

$$COE_t = \mu_0 + \mu_1 Cap_{t-1} + \mu_2 Controls_{t-1} + \epsilon_t$$

a. Where $COE_t$ = Cost of equity, $Cap_{t-1}$ = Equity, Tier 1 and CAR as a measure of bank capital, Controls = a bunch of bank and country level control variables. The parameter $\mu_1$ are expected to be negatively related with the bank's cost of equity. The logarithm of ratio variables can explain the dependent variable much better than the ratio itself (Lien et al. 2017). We have taken the logarithm of Gross Domestic Product, Asset Size and Deposit. This logarithmic application reduces the inconsistency of the variables and assists in neutralizing the size impact.

b. **Data**

We conduct the study based on secondary data collected from different sources like Bangladesh Bank, Bangladesh Bureau of Statistics publications, Dhaka Stock Exchange, Bangladesh Securities and Exchange Commission and World Bank database. For our research, we have collected data of 28 publicly listed banking companies of the period ranging from 2013 to 2019.

### c. The Implied Cost of Equity Capital

We estimate the dependent variable, the implied cost of equity (COE), as the average estimate obtained from the Dividend Discount Model (DDM) in line with the methods of Myron J. Gordon, 1956 (According to the DDM, the current price of a share of company is the discounted (discounted by Cost of Equity) present value of its future dividends.

$$P_t = \sum_{t=0}^{\infty} \frac{E_t D_t}{(1 + R_e)^t}$$

The detailed explanation of Dividend Discount Model we have used here to determine the cost of equity has been described in Annexure B.

### d. Bank Capital Variables

Bank capital is the primary test variable of our research. We have used three alternative measures of bank capital. These are: Financial Leverage (Equity), Tier 1 Capital ratio (Tier 1 Capital divided by Risk-weighted Asset) and CAR (Sum of Tier 1 and Tier 2 capital divided by Risk-weighted asset).

Financial leverage is the primary measure of capital in our research as equity holders have faith in this measure. They rely on this measure since this requires simple calculation and it represents the amount of high-quality capital. A high-quality capital ensures the highest loss absorption capacity. This measure also ignores the drawbacks that risk-weighted capital ratios entail.

Our second capital measure Tier 1 regulatory capital ratio is calculated by dividing Tier 1 capital by Risk-Weighted Asset. Our third capital measure is CAR (Capital Adequacy Ratio), which we have measured by dividing the sum of Tier 1 and Tier 2 capital by Risk-Weighted Assets. Though there may be flaws, the equity investors may rely on these measures of capital to assess the financial risk and to measure the required rate of return. We can mention here that Tier 1 regulatory capital includes the core capital of a bank, which comprises of common stock, retained earnings and perpetual non-cumulative preferred stock. In contrast, Tier 2 capital consists of the mix of debt and equity, subordinated debt, revaluation reserves and loan loss reserves.

### e. Control Variables

We incorporate some bank and macroeconomic specific variables to consider the factors other than the bank capital, which may impact the banks' cost of equity. Different bank-level factors other than the bank capital may influence the perception of equity investors for which they may require higher or lower risk premium. In our analysis, we control for variables which are- Inefficiency (INEFF), Return on Asset (ROA), Natural Log of Total Deposit (LNDEPOSIT),

Natural Log of Size (LNSIZE), Asset Quality (ASSTQULTY), and Natural Log of Gross Domestic Product (LNGDP)..

A high Non-Performing Loan implies that the bank's asset quality is not up to the mark. The equity investors generally demand a high required rate of return (the cost of equity) from a bank holding a large amount of loan loss provision because of loan repayment risk. We have also controlled the bank's management quality which we have labeled as inefficiency. We have measured the quality of bank's management by the ratio of salaries and benefits to total assets. It is generally expected that a high cost associated with salaries and benefits leads to high inefficiency. The more inefficient a bank is, the equity investors demand the high cost of equity. Equity holders also look into the bank earnings, which affect the cost of equity demanded by them noticeably.

To account for this, we control the variable- the return on assets calculated as the ratio of deposits to total assets- as it affects the cost of equity. The amount of deposit a bank signifies the health of bank. A bank holding a high amount of deposit enjoys more stability in its funding structure, which reduces its susceptibility to liquidity problems (Beltratti and Stulz, 2012; Berger and Bowman, 2013). Thus, high amount of deposit also reduces the cost of equity demanded by the equity holders. Our final bank-level variable is the natural logarithm of assets (Size). Equity investors consider that the larger a bank, the less risky it is due to better asset diversification (Demsetz and Strahan, 1997) and better monitoring of regulatory body. Additionally, equity investors perceive the larger banks too big to fail (Deng et al. 2007; Belkhir, 2013) and thus demand lower required rate of returns in the form of cost of capital compared to the small size banks.

In our research paper, we have also used the Natural Log of GDP as a control variable like various other studies (e.g., Belkhir et al., 2019; Chen et al., 2011; Chen et al., 2009; Hail and Leuz, 2006).

f. **Summary Statistics**

**Table2: Summary Statistics**

| Variable | Obs | Mean | Std. Dev. | Min | Max |
|---|---|---|---|---|---|
| COE | 196 | 0.150 | 0.036 | 0.070 | 0.250 |
| EQUITY | 196 | 0.095 | 0.015 | 0.048 | 0.137 |
| TIER1 | 196 | 0.069 | 0.015 | 0.032 | 0.115 |
| CAR | 196 | 0.126 | 0.015 | 0.100 | 0.180 |
| INEFF | 196 | 0.011 | 0.003 | 0.002 | 0.030 |
| ROA | 196 | 0.009 | 0.003 | 0.000 | 0.020 |
| LNDEPOSIT | 196 | 9.785 | 0.429 | 8.828 | 11.450 |
| LNSIZE | 196 | 10.071 | 0.428 | 9.098 | 11.645 |
| ASSTQULTY | 196 | 0.049 | 0.016 | 0.020 | 0.130 |
| LNGDP | 196 | 5.383 | 0.234 | 5.010 | 5.712 |

Table 02 shows the descriptive statistics of all the variables used in the model. There are 196 observations (28 publicly listed banking companies ranging from 2013 to 2019) for each variable. The mean cost of equity is 15% with a maximum of 25% and minimum of 7% only. The dataset shows that the volatility of cost of equity is 3.6%. The ratio of equity capital to total assets is 9.5% with a maximum of 13.7% and a minimum of 4.8%. In this case, volatility of equity capital to total asset ratio is 1.5% only. Lower level of volatility indicates the stability of the banks' equity though there is some higher volatility on the cost of equity. The ratio of Tier 1/Risk weighted assets (TIER1) and salaries & benefits to total asset (INEFF) is 6.9% and 12.6% respectively whereas the mean ROA is 0.9% with a maximum of 2% and a minimum of 0%. The ratio of logarithm of total deposit to total asset (LNDEPOSIT) and logarithm of GDP (LNGDP) is 9.78 and 5.38 respectively. Of all 196 observations, LNDEPOSIT, LNGDP and COE are comparatively more volatile than other variables. The mean ratio of total deposit to total asset is higher compared to other variables. Different features of our sample collected from publicly listed banks have different characteristics in terms of their capitalization, size, asset quality, profitability and liquidity which have resulted in the variation of standard deviation, maximum and minimum value of each variable. This may also happen due to the missing observations as there is a small sample of publicly listed banks.

## 4. Empirical Findings

The primary evidence of the relationship between the cost of equity and bank capital is shown in Table 3. In the table we show total six different regression outputs. While column 1 and 2 represents the findings of our estimations based on the EQUITY (equity to asset ratio) as a measure of bank capital, column 3 and 4 use TIER1 (Tier1 capital/Risk-Weighted Assets) and column 5 and 6 use CAR (total capital/Risk-Weighted Assets). Using the data of 28 publicly listed scheduled commercial banks of Bangladesh for the period 2013 to 2019 and employing fixed effect panel estimation with Driscoll Kraay standard errors, we observe that if the capital increases by one standard deviation, the cost of equity of banks (standard deviation 0.036 see table 2) drops by a (-0.439*0.036= -0.0158) 158 basis points holding all other variables remain constant. In the same way, a 10-percent increase in the level of capital would reduce the cost of equity by 4.39 percent. Consistent with the theory, the paper identifies a significant negative empirical relationship between the cost of equity and banks' capital maintenance level. In other words, the paper finds that banks with higher capital (equity to asset ratio) experience reduced cost of equity while the relationship is statistically significant and economically meaningful.

In column 2 of table 3, we check the robustness of the findings of column 1 by adding macroeconomic variable—the natural logarithm of GDP—in the regression model. However, the result depicts no change in the result of cost of equity —neither the statistical nor the economic significance—for inserting this macro variable unless slight increase of the coefficient of determinant (r square increased to 39.36 percent). The coefficient of EQUITY in column 2 is almost similar to the coefficient in model 1. To be specific, a 10-percent increase in the level of capital would reduce the cost of equity by about 5 percent. Regarding the control variables, we find the relationship is consistent with the literature. For example, the coefficient of INEFF is positive which means banks with high Non-Performing Loans increase the cost of equity as it decreases the loanable funds and deteriorates the quality of loan portfolio.

**Table 3: Output of the Regressions**

| Variables | 1 COE | 2 COE | 3 COE | 4 COE | 5 COE | 6 COE |
|---|---|---|---|---|---|---|
| Equity | -0.439*** | -0.503*** | | | | |
|  | (-3.59) | (-3.43) | | | | |
| TIER1 | | | -0.476** | -0.417** | | |
|  | | | (-2.86) | (-2.8) | | |
| CAR | | | | | -0.256*** | -0.154** |
|  | | | | | (-3.37) | (-2.4) |
| INEFF | 0.266** | 0.243** | 0.171 | 0.161 | 0.241 | 0.203 |
|  | (-0.69) | (-0.61) | (-0.59) | (-0.53) | (-0.61) | (-0.52) |
| ROA | 0.059*** | 0.061*** | 0.062*** | 0.061*** | 0.050*** | 0.052*** |
|  | (-19.77) | (-21.57) | (-19.09) | (-19.2) | (-14.15) | (-16.94) |
| LNDEPOSIT | -0.005** | -0.004* | -0.006*** | -0.006** | -0.004 | -0.003 |
|  | (-2.6) | (-1.82) | (-3.56) | (-2.49) | (-1.13) | (-0.98) |
| LNSIZE | 0.006 | -0.004 | 0.003 | -0.003 | 0.006** | -0.001 |
|  | (-1.81) | (-0.67) | (-0.93) | (-0.62) | (-2.94) | (-0.13) |
| ASSTQULTY | -0.195** | -0.158* | -0.166* | -0.158* | -0.259*** | -0.235*** |
|  | (-2.39) | (-1.95) | (-2.07) | (-2.12) | (-4.65) | (-4) |
| LNGDP | | 0.033** | | 0.022** | | 0.024** |
|  | | (-2.72) | | (-2.6) | | (-2.7) |
| Constant | 0.094 | 0.019 | 0.128* | 0.062* | 0.019 | -0.029 |
|  | (-1.54) | (-0.55) | (-2.28) | (-1.75) | (-0.38) | (-0.87) |
| Observations | 196 | 196 | 196 | 196 | 196 | 196 |
| R-Squared | 0.359 | 0.393 | 0.355 | 0.370 | 0.334 | 0.349 |

This table is the output of cross-sectional regressions derived from the equation.
*, ** and *** respectively mean 10%, 5% and 1% level of significance.

Similarly, the negative coefficients of ROA and LNDEPOSIT signify that the more profitable and liquid banks have lower cost. The size of the banks also has a negative impact on the banks' COE. The output suggests that the COE also depends on the household income levels—as indicated by the positive and significant coefficient for LNGDP.

We also test the relationship of COE with the level of quality capital. In column 3-4 we test the relationship with the TIER1 capital, which consists of broadly paid-up capital and retained earnings. The estimate shows that like the findings with EQUITY level, there is a significant negative relationship between COE and TIER1. The coefficient estimate on CAR is negative and significant at the 1 percent level and has a smaller magnitude compared to TIER1. Interestingly, while the TIER1 and CAR remain significant, the economic significance of CAR decreases (coefficient of TIER1 is 0.47 while the coefficient of CAR is 0.25). These findings imply that the COE is more directly related to the quality of capital. Investors apparently don't price the TIER2 capital, and the perceived financial risk is heavily priced by the core capital (TIER 1). The findings also suggest that COE declines when we employ more core capital. The finding is also in line with the theory that when banks have higher capital it reduces the risk of investors and depositors, which causes its stakeholders to require a lower cost of equity.

## 5. Conclusion and Policy Implications

This paper identifies the empirical relationship between the cost of equity and the banks' capital. Consistent with the theory, the paper identifies a significant negative empirical relationship between the cost of equity and banks' capital maintenance level. In other words, the paper finds that banks with higher capital experience reduced cost of equity and the relationship is statistically significant and economically meaningful. Using the data of 28 publicly listed scheduled commercial banks of Bangladesh for the period 2013 to 2019 and employing fixed-effect panel estimation with Driscoll Kraay standard, we observe that if the banks' level of equity increases by one standard deviation, the costs of equity of banks drops by 158 basis points holding all other variables constant. In the same way, a 10-percent increase in the level of capital would reduce the cost of equity by 4.39 percent. Our result also suggests that the relationship between cost of equity and the capital is heavily tilted towards the quality of capital—meaning the better the quality of capital, the more robust the relationship.

The finding of the paper has important policy implications as Bangladesh is implementing the Basel III capital framework. As the bankers often argue that raising the level of capital increases the cost of equity, this paper provides empirical evidence that the rising of capital instead reduces cost of equity. As capital minimizes the probability of bank failure, ensures sustainability, and protects the depositors in times of failure, banks should enhance capital maintenance and improve the quality of capital. The regulatory bodies could come forward to increase the systemic capital maintenance—by increasing the minimum capital requirement (Tier 1 Capital)—to ensure the sustainability of the financial system and reduce the probability of bank failure. Together, the results suggest that building on the progress made since the implementation of Basel III capital adequacy framework, some policy adjustment—such as raising the minimum requirement of Common Equity Tier 1 capital and Additional Tier 1 capital coupled with reducing the space of Tier 2 capital (relatively less quality capital)—would help further strengthen banks sustainability and financial sector stability.

30. Modigliani, F. and Miller, M.H., 1958. The cost of capital, corporation finance and the theory of investment. *The American economic review*, *48*(3), pp.261-297.

31. Naoaj, Md. Shah and Khan, Asif and Ahsan, Nazmul, The Emerging Asia Pacific Capital Markets: Bangladesh (March 18, 2021). CFA Institute Research Foundation Briefs, March 2021.

32. Rankin, E. and Idil, M.S., 2014. A Century of Stock-Bond Correlations| Bulletin–September Quarter 2014. *Bulletin*, (September).

33. Wu, X. and Wang, Z., 2005. Equity financing in a Myers–Majluf framework with private benefits of control. *Journal of Corporate Finance*, *11*(5), pp.915-945.
13

**Appendix A: Variables, Definitions and Sources**

| Variable | Definition | Source |
|---|---|---|
| **Panel A. Implied Cost of Equity** | | |
| COE | Cost of Equity is calculated as the average estimate obtained from the Dividend Discount Model (DDM) in line with the methods of Myron J. Gordon. | |
| **Panel B. Capital ratio variables** | | |
| EQUITY | Equity/ Total Assets ratio. | Bangladesh Bank |
| TIER1 | Tier 1 /Risk weighted assets ratio. | Bangladesh Bank |
| CAR | Eligible Capital/Risk-weighted assets ratio | Bangladesh Bank |
| **Panel C. Bank-level control variables** | | |
| INEFF | Salaries and benefits/total assets ratio. | Bangladesh Bank |
| ROA | The return on assets ratio. | Bangladesh Bank |
| LNDEPOSIT | Natural Logarithm of (Total deposits/Total assets) ratio. | Bangladesh Bank |
| LNSIZE | Natural logarithm of the Banks' Total Assets | Bangladesh Bank |
| ASSTQULTY | Total Non-Performing Loans/ Total Loans and Advance | Bangladesh Bank |
| **Panel D. Country-level control variables** | | |
| LNGDP | The logarithm of GDP in USD | World Bank Database |

**Appendix B: Cost of Equity Determination**

The investor decides to buy and hold shares based on the cost of equity (COE) measure. Two determining factors based on which cost of equity is generally measured are- Risk-Free Rate of Return, usually determined by the yield on long term government bond and Risk Premium that compensates the investors for buying and holding a risky asset. The cost of equity may rise in line with investors' perception on the economic growth outlook and the increase in uncertainty based on that outlook as in Rankin and Idil 2014. The cost of equity has to be proportionate to leverage as in Modigliani and Miller, 1958. Unfortunately, some factors cause the deviation of the cost of equity from what was predicted by Modigliani-Miller theorem. Because of the inclusion of risk tolerance and expectation in the cost of equity, it becomes unobservable. As a matter of fact, we can proxy the cost of equity based on the Dividend Discount Model as indicated in Gordon and Shapiro 1956.

$$R_e = \frac{D_{t+1}}{P_t + g}$$

Hereby $R_e$ is Cost of Equity, D is Dividend, P is Share Price and g is Expected Dividend Growth Rate. In order to compare with return on equity, we can replace Dividend with Earning multiplied by Dividend Payout Ratio.

$$R_e = \frac{\beta E_{t+1}}{P_t} + g$$

Here $\beta$ is the Dividend Payout Ratio and E is Earning. Based on the investor's perception, if dividend does not grow, then return on equity will be equivalent to the forward earnings yield determined by $(\frac{\beta E_{t+1}}{P_t})$. Return of equity will be equal to the forward earnings yield if a firm distributes all earnings as dividends. We can use the Return on Equity (ROE) as a proxy for the Cost of Equity if the cost of equity is unobservable. Taking ROE as a proxy for the cost of equity may mislead the outcomes as the former is measured by book value of equity and the market value of equity measures the latter. As ROE is measured only from current profits and COE is measured based on future dividend growth expectations, the outcome may further be misleading. As per the Dividend Discount Model, the fair value of a share can be measured based on the expected value of future stream of dividends which are then discounted back to present value using the cost of equity.

$$P_t = \sum_{t=0}^{\infty} \frac{E_t D_t}{(1 + R_e)^t}$$

Hereby Et represents the expectation operator. If a firm continues growing steadily at a rate of g, this model can be simplified to as follows as per Gordon and Shapiro 1956:

$$P_t = \frac{D_{t+1}}{R_e - g}$$

We can rearrange this equation to find out the cost of equity as follows:

$$R_e = \frac{D_{t+1}}{P_t} + g$$

As growth derives only from reinvesting retained earnings, the elimination of unobservable steady-state growth rate is possible and we can rearrange the equation as follows:

$$R_e = \frac{\beta E_{t+1}}{P_t} + g$$

$$= \frac{\beta E_{t+1}}{P_t} + (1-\beta)\frac{E_t}{B_t}$$

$$\approx \frac{E}{P} + (1-\beta)(\frac{E}{B} - \frac{E}{P})$$

Herby, the second line of the equation is $g = (1-\beta)\frac{E_t}{B_t}$, Earnings Yield (E/P) and ROE is $E/B = \pi_e$. If a firm does not grow (g=0), we can simplify the equation and write the ratio of ROE to COE (P/B ratio) as follows:

$$R_e = \frac{E}{P} \quad \text{Where g=0}$$

$$\therefore \frac{\pi_e}{R_e} = \frac{P}{B}$$

Fortunately, if we expect that dividend will grow (Growth is not zero), this relationship still holds. In this case, the ratio of ROE to COE is scaled by the weighted average P/B ratio. For bank that distributes at least half of its earnings as dividends and grows moderately, this equation converges the ratio of ROE to COE to P/B ratio. However, for bank that does not distribute dividend, ROE is always equal to COE. The reason behind this is that if a firm's capital proliferates, there remains no difference between the book value of equity and the market value of equity.

$$R_e = \frac{\beta E_{t+1}}{P_t} + g$$

Since $\pi_e = \frac{E}{B}$

$$\frac{\pi_e}{R_e} \approx \frac{E}{B} \times \frac{P}{\beta E + Pg}$$

$$\approx \frac{P}{B} \times \frac{E}{E\left[\beta + (1-\beta)\frac{P}{B}\right]}$$

$$\approx \frac{P}{B} \times \varepsilon^{-1} \quad \text{Where } \varepsilon^{-1} = \beta + (1-\beta)\frac{P}{B}$$